\documentclass[prd,twocolumn,amsmath,nofootinbib,amssymb]{revtex4}
\usepackage{graphicx,color,dcolumn,booktabs,bm}
\usepackage{longtable,lscape}
\usepackage{pdfpages}
\usepackage{txfonts}
\usepackage{overpic}
\usepackage{amssymb}
\usepackage{makecell}
\usepackage{indentfirst}
\usepackage{feynmf}   
\usepackage{slashed}  
\usepackage{cases}
\usepackage{color}
\usepackage{multirow}
\usepackage{threeparttable}
\usepackage{epstopdf}
\usepackage{enumerate}
\usepackage{subfigure}
\usepackage{diagbox}
\usepackage{graphicx,color,dcolumn,booktabs,bm}
\usepackage{mathrsfs}
\usepackage{cancel}
\usepackage{float}
\usepackage[misc]{ifsym}
\usepackage[colorlinks,
            citecolor=blue,
            anchorcolor=red,
            menucolor=red,
            linkcolor=red,
            filecolor=red,
            runcolor=red,
            urlcolor=blue,
            frenchlinks=red]{hyperref}
\usepackage{array}
\usepackage{booktabs}

\begin{document}


\title{Comparing relativistic and non-relativistic quark pair creation models}

\author{Xiu-Li Gao}
\affiliation{School of Physics, Southeast University, Nanjing 211189,
P.~R.~China}

\author{Yu-Hui Zhou}
\affiliation{School of Physics, Southeast University, Nanjing 211189,
P.~R.~China}
\author{Bin Wu}
\affiliation{School of Physics, Southeast University, Nanjing 211189,
P.~R.~China}
\author{Zhi-Yong Zhou}
\email[]{Corresponding author: zhouzhy@seu.edu.cn}

\affiliation{School of Physics, Southeast University, Nanjing 211189,
P.~R.~China}

%

\begin{abstract}
We investigate the strong decay properties of light unflavored and strange mesons within a relativistic quark-pair-creation (QPC) framework, and compare the results with those obtained in the conventional non-relativistic QPC model. Our analysis shows that, within the present theoretical and experimental uncertainties, the relativistic QPC model yields predictions for strong decay widths of comparable overall quality to those of the non-relativistic QPC model. This indicates that the non-relativistic QPC approach remains adequate for estimating decay widths in most practical applications. Nevertheless, owing to the inclusion of Lorentz boosts and Wigner rotations, the relativistic QPC model exhibits a stronger suppression of decay amplitudes in the high-energy region. This feature may be useful in studies based on unquenched quark models, where the relativistic QPC coupling could lead to more controlled meson-loop effects and mass shifts.

\end{abstract}


\maketitle

\section{Introduction}
Strong hadronic decays provide an important window on the nonperturbative dynamics of Quantum Chromodynamics (QCD). Since these processes typically occur at energy scales where perturbative methods are not applicable, a number of phenomenological approaches have been developed to describe transitions between hadronic states. Among them, the quark-pair-creation (QPC) model, more commonly known as the $^{3}P_{0}$ model, has long been used as a standard framework for Okubo-Zweig-Iizuka~(OZI)-allowed strong decays of mesons and baryons~\cite{Micu:1968mk,LeYaouanc:1972vsx,LeYaouanc:1973ldf,Roberts:1992esl,Capstick:1993kb,Capstick:2000qj,Blundell:1995ev,Barnes:1996ff}.

In the QPC model, the decay is assumed to proceed through the creation of an additional quark-antiquark pair from the vacuum with quantum numbers $J^{PC}=0^{++}$. In spectroscopic notation, this corresponds to a spin-triplet $P$-wave $q\bar q$ pair, which gives the model its name. The created pair then recombines with the constituents of the parent hadron to form the final-state hadrons. Although phenomenological in nature, this construction has been found to provide a useful description of a broad class of OZI-allowed transitions.

The model was first proposed by Micu~\cite{Micu:1968mk} and was subsequently developed in a systematic way by Le Yaouanc and collaborators, who formulated it in terms of explicit spin, flavor, and spatial wave functions. As constituent quark models became more widely used, the QPC framework was adopted extensively in studies of conventional mesons and baryons, owing both to its relative simplicity and to its reasonable phenomenological performance in many applications~\cite{Capstick:2000qj,Capstick:1993kb,Blundell:1995ev,Song:2015nia,Lu:2006ry,Zhang:2006yj,Feng:2022hwq,Li:2022ybj,Song:2014mha,Chen:2016iua,Li:2020xzs,Pang:2019ttv,Pang:2018gcn,Xiao:2018iez,Ferretti:2015rsa,Sun:2009tg,Garcia-Tecocoatzi:2022zrf}.

At the same time, considerable effort has been devoted to understanding the microscopic origin of the $^{3}P_{0}$ vertex. One frequently discussed interpretation relates the pair-creation operator to the Lorentz structure of confinement: if the confining interaction contains a dominant scalar component, then pair creation with vacuum quantum numbers $0^{++}$ can arise rather naturally. Related analyses have also attempted to connect the phenomenological decay operator with more microscopic interquark interactions, often through suitable non-relativistic reductions of the underlying dynamics~\cite{Ackleh:1996yt,Kokoski:1985is}.

The QPC model has also been widely applied in heavy-quark spectroscopy, where it is commonly used to estimate open-flavor strong decay widths of charmonium, bottomonium, and heavy-light mesons. In such systems, the non-relativistic approximation is often expected to be more reliable. For light hadrons and highly excited states, however, its applicability is less clear, since larger internal momenta, stronger recoil effects, and relativistic spin couplings may become more important. In its conventional implementation, the QPC model is essentially non-relativistic: hadron wave functions are usually taken from non-relativistic or semirelativistic quark models, and the decay operator is typically evaluated with non-relativistic kinematics. It is therefore worthwhile to examine whether a relativistic treatment may affect phenomenological predictions in these regimes, which is still absent in the literature.

Beyond the calculation of isolated partial widths, the QPC mechanism has also been used extensively in unquenched quark models, where it couples valence quark-model states to meson-meson continuum channels and thereby generates self-energy corrections, configuration mixing, and continuum components in hadron wave functions~\cite{Heikkila:1983wd,Kalashnikova:2005ui,Zhou:2017dwj}. In such applications, not only the on-shell decay widths but also the off-shell energy dependence of the vertex functions may become relevant, which makes the role of relativistic effects of particular interest.

In this work, we present an exploratory comparison between the conventional non-relativistic QPC model and a relativistic formulation developed along the lines of Refs.~\cite{Fuda:2012xd,Zhou:2020vnz}. Using meson wave functions from the Godfrey-Isgur (GI) model, we calculate partial decay widths for a representative set of meson states in both approaches. We find that the two formulations generally give similar results for many channels, while the relativistic treatment appears to provide some improvement for certain highly excited states. We also compare the corresponding vertex functions as functions of the center-of-mass energy and find that the relativistic formulation exhibits a noticeably milder high-energy behavior. This feature may be relevant in applications where the energy dependence of the hadronic vertices plays a significant role.

This paper is organized as follows. In Sec.~\ref{section2}, we briefly review the non-relativistic QPC model in Sec.~\ref{seciton21} and the relativistic QPC model in Sec.~\ref{section22}. Numerical results for the strong decays of light and strange mesons are presented in Sec.~\ref{section3}, where they are compared with available experimental data and with the predictions of the non-relativistic model. Finally, Sec.~\ref{section4} contains our conclusions.

\section{The QPC model}\label{section2}

\subsection{The non-relativistic version}\label{seciton21}

In the conventional QPC model~\cite{Micu:1968mk,Blundell:1995ev, Ackleh:1996yt}, a quark-antiquark pair with the quantum number,  $J^{PC}=0^{++}$, is created from the vacuum and then they are combined with the quark and antiquark in the initial hadron state to form the final states. For the meson case, the wavefunction of a mock meson $A$ is defined by
\begin{align}
& \left|A\left(n_A{ }^{2 S_A+1} L_{A J_A M_{J_A}}\right)\left(\mathbf{P}_A\right)\right\rangle \notag \\ 
\equiv & \sum_{M_{L_A}, M_{S_A}}\left\langle L_A M_{L_A} S_A M_{S_A} \mid J_A M_{J_A}\right\rangle \notag\\
& \times \int \mathrm{d}^3 \mathbf{p}_A \psi^A_{L_A M_{L_A}}\left(\mathbf{p}_A\right) \chi_{{S_A}M_{S_A}}^{12} \phi_A^{12} \omega_A^{12} \notag\\
& \times\left|q_1\left(\frac{\mu_1}{\mu_1+\mu_2} \mathbf{P}_A+\mathbf{p}_A\right) \bar{q}_2\left(\frac{\mu_2}{\mu_1+\mu_2} \mathbf{P}_A-\mathbf{p}_A\right)\right\rangle.\label{mockstate1}
\end{align}
In this representation, the  c.m. momentum of meson $A$ denotes $\mathbf{P}_A$, and the relative momentum of the quark-antiquark pair $\mathbf{p}_A$ is integrated over all values.   $\left|S_A, M_{S_A}\right\rangle$, $\left|L_A, M_{L_A}\right\rangle$, and $\left|J_A, M_{J_A}\right\rangle$ are respectively spin, relative angular momentum, and total angular momentum with their third components. $\chi_{S_A M_{S_A}}^{12}, \phi_A^{12}$ and $\omega_A^{12}$ are spin, flavor and color wave funtions of the quark pair respectively. $\psi^A_{L_A M_{L_A}}\left(\mathbf{p}_A\right)$ is the spatial wave function of c.m. frame in momentum-space; $\mu_1$ and $\mu_2$ are the masses of the quark and antiquark, and  $\left|q_1\left(\mathbf{p}_1\right) \bar{q}_2\left(\mathbf{p}_2\right)\right\rangle$ is the plane-wave state of a free quark and antiquark. 
Normalizations of the mock meson could be chosen as
\begin{align}
\left\langle A(\mathbf{P}_i) \mid A\left(\mathbf{P}_j\right)\right\rangle&= \delta^3\left(\mathbf{P}_i-\mathbf{P}_j\right).\label{normalization1}
\end{align}
Notice that the normalization is chosen for convenience of the later comparison with the relativistic version.

 For the meson decay $A\rightarrow BC$, with quark labels denoted as $(12)\rightarrow(14)+(32)$, the $S$-matrix reads
\begin{align}
&\langle BC|S|A\rangle = I-2 \pi i \delta\left(E_f-E_i\right) \langle BC|T|A\rangle\nonumber\\
&= I-2 \pi i \delta\left(E_f-E_i\right) \delta^3\left(\mathbf{P}_B+\mathbf{P}_C-\mathbf{P}_A\right) \mathcal{M}^{M_{J_A} M_{J_B} M_{J_C}},
\label{Smatrix}
\end{align}
with the transition operator $T_{_{NR}}$  expressed as

\begin{align}
T_{_{NR}}=&-3 \gamma_{_{NR}} \sum_m\langle 1 m 1-m \mid 00\rangle \int \mathrm{d}^3 \mathbf{p}_3 \mathrm{~d}^3 \mathbf{p}_4 \delta^3\left(\mathbf{p}_3+\mathbf{p}_4\right)\notag\\
&\times \mathcal{Y}_1^m\left(\frac{\mathbf{p}_3-\mathbf{p}_4}{2}\right) \chi_{1-m}^{34} \phi_0^{34} \omega_0^{34} b_3^{\dagger}\left(\mathbf{p}_3\right) d_4^{\dagger}\left(\mathbf{p}_4\right),\label{Tmatrix1}
\end{align}
where $\gamma_{_{NR}}$ represents the production strength parameter. $b_3^{\dagger}\left(\mathbf{p}_3\right)$ and $d_4^{\dagger}\left(\mathbf{p}_4\right)$ are the creation operators of quark and antiquark. $\chi_{1-m}^{34}$, $\phi_0^{34}$, $\omega_0^{34}$ represents spin triplet, flavor singlet and color singlet wave functions. $\mathcal{Y}_l^m(\mathbf{p})=p^lY_l^m(\theta_p,\phi_p)$ is a solid harmonic that gives the momentum-space distribution of the created pair. All these elements are used to ensure the created quark and antiquark to possess the vacuum quantum numbers.

The helicity amplitude could be obtained by standard derivations as
\begin{align}
& \mathcal{M}^{M_{J_A} M_{J_B} M_{J_C}}\notag\\
&=  \gamma_{NR} \sum_{\substack{
M_{L_A}, M_{S_A}, M_{L_B}, M_{S_B} \\
M_{L_C}, M_{S_C}, m
}}\left\langle L_A M_{L_A} S_A M_{S_A} \mid J_A M_{J_A}\right\rangle \notag\\
& \times\left\langle L_B M_{L_B} S_B M_{S_B} \mid J_B M_{J_B}\right\rangle\left\langle L_C M_{L_C} S_C M_{S_C} \mid J_C M_{J_C}\right\rangle \notag\\
& \times\langle 1 m 1-m \mid 00\rangle\left\langle\chi_{S_B M_{S_B}}^{14} \chi_{S_C M_{S_C}}^{32} \mid \chi_{S_A M_{S_A}}^{12} \chi_{1-m}^{34}\right\rangle \notag\\
& \times\left[\left\langle\phi_B^{14} \phi_C^{32} \mid \phi_A^{12} \phi_0^{34}\right\rangle I_{M_{L_BL_B}}^{M_{L_A}m}\left(\mathbf{q}\right)\right. \notag\\
& \left.+(-1)^{1+S_A+S_B+S_C}\left\langle\phi_B^{32} \phi_C^{14} \mid \phi_A^{12} \phi_0^{34}\right\rangle I_{M_{L_BL_B}}^{M_{L_A}m}\left(-\mathbf{q}\right)\right],\label{NRHelicity}
\end{align}
where $\left\langle\chi_{S_B M_{S_B}}^{14}\chi_{S_C M_{S_C}}^{32} \mid \chi_{S_A M_{S_A}}^{12} \chi_{1-m}^{34}\right\rangle$ is the spin matrix element, the color matrix element $\left\langle\omega_B^{14} \omega_C^{32} \mid \omega_A^{12} \omega_0^{34}\right\rangle=1/3$, cancelling the factor 3 in the definition of $T$ operator. The flavor matrix element $\left\langle\phi_B^{14} \phi_C^{32} \mid \phi_A^{12} \phi_0^{34}\right\rangle=1/\sqrt{3}$ with the assumption of flavor $SU(3)$ symmetry. The space integral factor is given by
\begin{align}
I_{M_{L_BL_B}}^{M_{L_A}m}\left(\mathbf{q}\right)= & \int \mathrm{d}^3 \mathbf{p} ~\psi^{B*}_{L_B M_{L_B}}\left(\frac{m_3}{m_1+m_3} \mathbf{q}+\mathbf{p}\right)\notag\\
& \times \psi^{C*}_{L_C M_{L_C}}\left(\frac{m_3}{m_2+m_3} \mathbf{q}+\mathbf{p}\right) \notag\\
& \times \psi^A_{L_A M_{L_A}}(\mathbf{q}+\mathbf{p}) \mathcal{Y}_1^m(\mathbf{p}),\label{amplitudeM}
\end{align}
where we have taken $\mathbf{q}=\mathbf{P}_B=-\mathbf{P}_C$, $\mathbf{p}=\mathbf{p}_3$.

By choosing the momentum $\mathbf{q}$ along the $z$-axis and  applying the Jacob-Wick formula \cite{Jacob:1959at}, the partial wave amplitude is expressed as
\begin{align}
 &M^{S L}(q)\notag\\
 =&\frac{\sqrt{4 \pi(2 L+1)}}{2 J_A+1} \sum_{M_{J_B}, M_{J_C}}\left\langle L 0 S\left(M_{J_B}+M_{J_C}\right) \mid J_A\left(M_{J_B}+M_{J_C}\right)\right\rangle \notag\\
& \times\left\langle J_B M_{J_B} J_C M_{J_C} \mid S\left(M_{J_B}+M_{J_C}\right)\right\rangle  M^{(M_{J_A}=M_{J_B}+M_{J_C}) M_{J_B} M_{J_C}}(q \hat{z}),\label{JacobWick}
\end{align}
where $S$ denotes total spin and $L$ denotes the relative angular momentum of final mesons.
Thus, the partial decay width of meson $A$ could be calculated through the following formula
\begin{align}\label{decaywidth}
\Gamma^{S L}=2\pi\frac{qE_BE_Cs}{M_A}\left|M^{S L}\right|^2,
\end{align}
where $E_B$ and $E_C$ are energies of $B$ and $C$ in the c.m. frame of $A$ and $s\equiv 1/(1+\delta_{BC})$ is statistical factor when $B$ and $C$ are identical.

\subsection{The relativistic quark pair creation model}\label{section22}

In the non-relativistic QPC model, the relative wave function of the quark-antiquark pair in a meson is defined in the meson rest frame, and Galilean transformations are implicitly assumed in the construction of the decay amplitude. Such a treatment is generally adequate only when the final-state momenta are sufficiently small. For decays involving sizable recoil, however, a proper Lorentz-covariant treatment becomes necessary~\cite{Fuda:2012xd,Zhou:2020vnz}. In the following, we denote a four-momentum by $p=(p^{0},\mathbf{p})$, and use $l_{c}(p)$ to represent the canonical Lorentz boost from the original inertial frame to a frame moving with velocity $\mathbf{v}=-\mathbf{p}/p^{0}$.

The relativistic mock state of meson $A$ with momentum $\mathbf{p}$ in the lab frame is constructed by applying canonical Lorentz boosts $l_c(p)$ to the quark and antiquark state
\begin{align}
& \big|A({}^{2s_A+1}l_{j_A m_A})(\mathbf{p})\big\rangle \notag\\
= &\sum_{m_l,m_s,m_1,m_2m_1',m_2'}
\int d^3 \mathbf{k}\, \psi^A_{l_A m_A}(\mathbf{k})\,
|\mathbf{p}_1, s_1 m_1'\rangle \otimes |\mathbf{p}_2, s_2 m_2'\rangle \,
\phi_A^{12}\, \omega_A^{12} \notag\\
& \times \mathscr{D}_{m_1' m_1}^{s_1}[r_c(l_c(p), k_1)]
\mathscr{D}_{m_2' m_2}^{s_2}[r_c(l_c(p), k_2)] \notag\\
& \times \langle s_1 s_2 m_1 m_2 | s_A m_{s_A}\rangle
\langle l_A s_A m_{l_A} m_{s_A} | j_A m_{j_A}\rangle \notag\\
& \times
\left(\frac{\varepsilon_1(\mathbf{p}_1)}{\varepsilon_1(\mathbf{k})}
\frac{\varepsilon_2(\mathbf{p}_2)}{\varepsilon_2(-\mathbf{k})}
\frac{W_{12}(\mathbf{k})}{E_{12}(\mathbf{p},\mathbf{k})}\right)^{1/2},
\label{mockstate}
\end{align}
where ${p}_1$ and ${p}_2$ denotes the four-momentum of  quark and antiquark in the meson in the lab frame and ${k}_1$ and ${k}_2$
their four-momentum in the c.m. frame of meson. $p\equiv (p^0,\mathbf{p})=p_1+p_2=(\varepsilon_1(\mathbf{p}_1+\varepsilon_2(\mathbf{p}_2),\mathbf{p}_1+\mathbf{p}_2) $ and 
\begin{align}
&\varepsilon_i(\mathbf{k})=(\mathbf{k}^2+\mu_i^2)^{1/2},\nonumber\\
&W_{12}(\mathbf{k})=\varepsilon_1(\mathbf{k})+\varepsilon_2(-\mathbf{k}),\nonumber\\
&E_{12}(\mathbf{p}, \mathbf{k})=({W_{12}(\mathbf{k})^2+\mathbf{p}^2})^{1/2}.
\end{align}
$r_c(a,p)\equiv l_c^{-1}(ap)al_c(p)$ is the Wigner rotation, $\mathscr{D}^{(s)}_{m' m}\bigl(R\bigr)$ is the  matrix representation of the rotation $R$. $\mathbf{k}$ is the relative momentum of quark in the c.m. frame of the meson. The factors are defined so as to obtain the  normalization condition 
\begin{align}
&\big\langle A({}^{2s_A+1}l_{j_A m_A})(\mathbf{p}') |A({}^{2s_A+1}l_{j_A m_A})(\mathbf{p})\big\rangle=\delta^3(\mathbf{p}-\mathbf{p}').
\label{normalization2}
\end{align}
Here $\psi^A_{lm}$ is the relative wave function of the $q\bar q$ pair in meson $A$ and its normalization satisfies $\int d^3k\, |\psi^A_{lm}(\mathbf{k})|^2 = 1$. The detailed derivation of two-particle state angular momentum representation could be found in appendix \ref{appendixA}.

 The interaction Hamiltonian is obtained from the quantum field theory 
\begin{align}
H_I=\gamma \int\mathrm{d}^3x\bar\psi(x)\psi(x),\ \ \  t=0,
\end{align}
where $\psi(x)$ denotes the Dirac field operator defined at the space-time point $x$ and an instant interaction is assumed. 
The transition operator $T_{_{RL}}$ is derived and  written down as~\cite{Fuda:2012xd}
\begin{align}
T_{_{RL}} =& - \sqrt{8 \pi}\, \gamma_{_{RL}} \int \frac{d^3 \mathbf{p}_3 d^3 \mathbf{p}_4}
{\sqrt{\varepsilon_3(\mathbf{p}_3)\,\varepsilon_4(\mathbf{p}_4)}}
\delta^{(3)}(\mathbf{p}_3+\mathbf{p}_4)\notag\\
& \times
\sum_{m,m_3,m_4}
\langle 1 m, 1{-}m | 00\rangle  \mathscr{Y}_1^m\!\left(\tfrac{\mathbf{p}_3-\mathbf{p}_4}{2}\right)
\langle \tfrac{1}{2} m_3, \tfrac{1}{2} m_4 | 1,{-}m \rangle\,\notag\\
& \times
\phi_0^{34}\, \omega_0^{34}\,
b_{m_3}^\dagger(\mathbf{p}_3)\, d_{m_4}^\dagger(\mathbf{p}_4),
\label{Tmatrix}
\end{align}
where the notations are similarly defined as in Eq.(\ref{Tmatrix1}). The helicity amplitude in the c.m.~frame of meson $A$ is obtained  as
\begin{widetext}
\begin{align}
& \mathscr{M}^{m_{j_A} m_{j_B} m_{j_C}}(\mathbf{q}) \notag\\
& =\sum_{m_{l_B} m_{s_B} m_{l_C} m_{s_C} m_{l_A} m_{s_A} m}\left\langle l_A s_A m_{l_A} m_{s_A} \mid j_A m_{j_A}\right\rangle \left\langle l_B s_B m_{l_B} m_{s_B} \mid j_B m_{j_B}\right\rangle\left\langle l_C s_C m_{l_C} m_{s_C} \mid j_C m_{j_C}\right\rangle\langle 1, m, 1,-m \mid 0,0\rangle \notag\\
& \times\{\left\langle\phi_B^{14} \phi_C^{32} \mid \phi_0^{34} \phi_A^{12}\right\rangle \int \mathrm{d}^3 \mathbf{k} \frac{(-\sqrt{8 \pi} \gamma_{RL} / 3)}{\varepsilon_3\left(\mathbf{p}_3\right)} \psi_{l_B m_{l_B}}^{B *}(\mathbf{k}) \psi_{l_C m_{l_C}}^{C *}\left(\mathbf{k}^{\prime}\right) \psi_{l_A m_{l_A}}^A\left(\mathbf{p}_1\right) \mathscr{Y}_1^m\left(\mathbf{p}_3\right) \notag\\
& \times \sum_{\substack{m_1 m_4 m_3 m_2
m_1^{\prime} m_4^{\prime} m_3^{\prime} m_2^{\prime}}}\left\langle s_1 s_2 m_1^{\prime} m_2^{\prime} \mid s_A m_{s_A}\right\rangle\left\langle s_1 s_4 m_1 m_4 \mid s_B m_{s_B}\right\rangle\left\langle s_3 s_2 m_3 m_2 \mid s_C m_{s_C}\right\rangle\left\langle s_3 s_4 m_3^{\prime} m_4^{\prime} \mid 1,-m\right\rangle \notag\\
& \times \mathscr{D}_{m_1^{\prime} m_1}^{(1 / 2) *}\left[r_c\left(l_c\left(q_1\right), k_1\right)\right]  \mathscr{D}_{m_4^{\prime} m_4}^{(1 / 2) *}\left[r_c\left(l_c\left(q_1\right), k_4\right)\right] \mathscr{D}_{m_3^{\prime} m_3}^{(1 / 2) *}\left[r_c\left(l_c\left(q_2\right), k_3\right)\right]  \mathscr{D}_{m_2^{\prime} m_2}^{(1 / 2) *}\left[r_c\left(l_c\left(q_2\right), k_2\right)\right] \notag\\
& \times\left(\frac{\varepsilon_1\left(\mathbf{p}_1\right)}{\varepsilon_1(\mathbf{k})} \frac{\varepsilon_4\left(\mathbf{p}_4\right)}{\varepsilon_4(\mathbf{k})} \frac{W_{14}(\mathbf{k})}{E_{14}(\mathbf{q}, \mathbf{k})}\right)^{1 / 2}\left(\frac{\varepsilon_3\left(\mathbf{k}^{\prime}\right)}{\varepsilon_3\left(\mathbf{p}_3\right)} \frac{\varepsilon_2\left(\mathbf{k}^{\prime}\right)}{\varepsilon_2\left(\mathbf{p}_2\right)} \frac{E_{32}\left(-\mathbf{q}, \mathbf{k}^{\prime}\right)}{W_{32}\left(\mathbf{k}^{\prime}\right)}\right)^{1 / 2}\notag\\
&+\left\langle\phi_B^{32} \phi_C^{14} \mid \phi_0^{34} \phi_A^{12}\right\rangle \int \mathrm{d}^3 \mathbf{k'} \frac{(-\sqrt{8 \pi} \gamma / 3)}{\varepsilon_3\left(\mathbf{p}_3\right)} \psi_{l_B m_{l_B}}^{B *}(\mathbf{k'}) \psi_{l_C m_{l_C}}^{C *}\left(\mathbf{k}\right) \psi_{l_A m_{l_A}}^A\left(\mathbf{p}_1\right) \mathscr{Y}_1^m\left(\mathbf{p}_3\right) \notag\\
& \times \sum_{\substack{m_1 m_4 m_3 m_2
m_1^{\prime} m_4^{\prime} m_3^{\prime} m_2^{\prime}}}\left\langle s_1 s_2 m_1^{\prime} m_2^{\prime} \mid s_A m_{s_A}\right\rangle\left\langle s_3 s_2 m_3 m_2 \mid s_B m_{s_B}\right\rangle\left\langle s_1 s_4 m_1 m_4 \mid s_C m_{s_C}\right\rangle\left\langle s_3 s_4 m_3^{\prime} m_4^{\prime} \mid 1,-m\right\rangle \notag\\
& \times \mathscr{D}_{m_3^{\prime} m_3}^{(1 / 2) *}\left[r_c\left(l_c\left(q_1\right), k_3\right)\right]  \mathscr{D}_{m_2^{\prime} m_2}^{(1 / 2) *}\left[r_c\left(l_c\left(q_1\right), k_2\right)\right] \mathscr{D}_{m_1^{\prime} m_1}^{(1 / 2) *}\left[r_c\left(l_c\left(q_2\right), k_1\right)\right]  \mathscr{D}_{m_4^{\prime} m_4}^{(1 / 2) *}\left[r_c\left(l_c\left(q_2\right), k_4\right)\right] \notag\\
& \times\left(\frac{\varepsilon_3\left(\mathbf{p}_3\right)}{\varepsilon_3(\mathbf{k'})} \frac{\varepsilon_2\left(\mathbf{p}_2\right)}{\varepsilon_2(\mathbf{k'})} \frac{W_{32}(\mathbf{k'})}{E_{32}(\mathbf{q}, \mathbf{k'})}\right)^{1 / 2}\left(\frac{\varepsilon_1\left(\mathbf{k}\right)}{\varepsilon_1\left(\mathbf{p}_1\right)} \frac{\varepsilon_4\left(\mathbf{k}\right)}{\varepsilon_4\left(\mathbf{p}_4\right)} \frac{E_{14}\left(-\mathbf{q}, \mathbf{k}\right)}{W_{14}\left(\mathbf{k}\right)}\right)^{1 / 2}\},\label{RLHelicity}
\end{align}
\end{widetext}
where the factor 1/3 comes from the overlap of the color wave functions. $\mathbf{k}$ is the three-momentum of quark 1 in the c.m. frame of 14 system, and $\mathbf{k}'$ is the three-momentum of antiquark 3 in the c.m. frame of 32 system. If quark 1 and antiquark 2 are of the same mass, $\mathbf{k}=\mathbf{k}'$, and the normalization factors will cancel. $\mathbf{p}_1$ is the three-momentum of the quark 1 in the c.m. frame of meson $A$, and $\mathbf{q}$ is the three-momentum of meson $B$ in the c.m. frame of $A$. $q_1$ and $q_2$ represent the four-momenta of meson $B$ and $C$ respectively. In this formula, there is also only one parameter $\gamma_{RL}$ to represent the production strength from the vacuum.

The transformation from helicity amplitudes to partial-wave amplitudes is carried out using the Jacob-Wick formula, in the same manner as in Eq.~(\ref{JacobWick}), after which the decay width is obtained from Eq.~(\ref{decaywidth}). It should be noted that, in the relativistic formulation [cf. Eq.~(\ref{RLHelicity})], the spatial and spin degrees of freedom are intertwined in the integration over the relative momentum $\mathbf{k}$ or $\mathbf{k}'$. Consequently, unlike in the non-relativistic case, these contributions cannot in general be factorized.

\begin{table*}[t]
\centering
\renewcommand\arraystretch{1.3}
\tabcolsep=0.3cm
\begin{tabular}{cllccc}
\toprule[1.6pt]
$J^{PC}(n^{2s+1}L_J)$ & Decay &  $\Gamma_{\text{exp}}$\cite{ParticleDataGroup:2024cfk} & $\Gamma_{NL}(\gamma_{NR}=18.3/14.7)$ & $\Gamma_{RL}(\gamma_{RL}=7.2/5.0)$   \\
\toprule[0.8pt]
$1^{--}(1^3S_1)$ & $\rho(770) \to \pi\pi$ & $147.4\pm 0.8$ & 147/95 & 147/70  \\
\cline{1-5}
$1^{--}(2^3S_1)$ & $\rho(1450) \to \omega\pi $  & $84.0\pm 12.6$ & 106/68 & 179/85  \\
\cline{1-5}
$1^{+-}(1^1P_1)$ & $b_1(1235) \to \omega\pi$  & $142.0\pm 9.0$ & 134/86 & 147/70 \\
\cline{1-5}
$1^{++}(1^3P_1)$ & $a_1(1260) \to \rho\pi $   & $234\pm 49$ & 457/293 & 500/238  \\
\cline{1-5}
$2^{++}(1^3P_2)$ & $a_2(1320) \to \rho\pi $  & $75.0\pm 4.5$ & 34/22 & 45/22  \\
$2^{++}(1^3P_2)$ & $a_2(1320) \to \pi\eta $  & $15.5\pm 1.5$ & 11/7 & 10/5  \\
$2^{++}(1^3P_2)$ & $a_2(1320) \to K\bar{K} $  & $5.2\pm 0.9$ & 4/2 & 4/2  \\
\cline{1-5}
$2^{++}(2^3P_2)$ & $a_2(1700) \to K\bar{K} $  & $4.9\pm 3.2$ & 0.9/0.6 & 0.5/0.2  \\
\cline{1-5}
$2^{-+}(1^1D_2)$ & $\pi_2(1670) \to f_2(1270)\pi $  & $145.3\pm 9.5$ & 131/84 & 137/66  \\
$2^{-+}(1^1D_2)$ & $\pi_2(1670) \to \rho\pi $  & $80.0\pm 10.7$ & 126/81 & 144/69  \\
$2^{-+}(1^1D_2)$ & $\pi_2(1670) \to K^*(892)\bar{K}$  & $10.8\pm 3.6$ & 25/16 & 43/21  \\
\cline{1-5}
$3^{--}(1^3D_3)$ & $\rho_3(1690) \to \pi\pi $   & $38.0\pm 3.2$ & 107/69 & 66/32  \\
$3^{--}(1^3D_3)$ & $\rho_3(1690) \to \omega\pi $  & $25.8\pm 9.8$ & 20/13 & 24/11  \\
$3^{--}(1^3D_3)$ & $\rho_3(1690) \to K\bar{K} $   & $2.5\pm 0.4$ & 2/1 & 5/2  \\
\toprule[1.6pt]
\end{tabular}
\caption{The strong decay widths of isovector mesons $\left(u\bar{d}, 1/\sqrt{2}(u\bar{u}-d\bar{d}), d\bar{u}\right)$ with two different $\gamma_{NR}$~(18.3 or 14.7)  and $\gamma_{RL}$~(7.2 or 5.0) are compared with those of candidates in PDG. The unit is MeV. }
\label{uu results}
\end{table*}

\begin{table*}[t]
\centering
\renewcommand\arraystretch{1.3}
\tabcolsep=0.3cm
\begin{tabular}{cllccc}
\toprule[1.6pt]
$J^{PC}(n^{2s+1}L_J)$ & Decay & $\Gamma_{\text{exp}}$\cite{ParticleDataGroup:2024cfk} & $\Gamma_{NL}(\gamma_{NR}=18.3/14.7)$ & $\Gamma_{RL}(\gamma_{RL}=7.2/5.0)$  \\
\toprule[0.8pt]
$1^{--}(1^3S_1) $ & $\phi(1020) \to K^+K^-$ & $2.12\pm 0.02$ & 2/1 & 3/2 \\
\cline{1-5}
$2^{++}(1^3P_2)$ & $f_2'(1525) \to K\bar{K} $  &  $63.4\pm 6.3$ & 130/84 & 126/60 \\
$2^{++}(1^3P_2)$ & $f_2'(1525) \to \eta\eta $  &  $7.5\pm 1.8$ & 24/15 & 20/10 \\
\cline{1-5}
$3^{--}(1^3D_3)$ & $\phi_3(1850) \to K\bar{K}$ & $31^{+10}_{-8}$ & 111/72 & 100/48 \\
$3^{--}(1^3D_3)$ & $\phi_3(1850) \to K^*(892)\bar{K}$ &  $56^{+18}_{-14}$ & 130/84& 184/88 \\
\toprule[1.6pt]
\end{tabular}
\caption{The strong decay widths of isoscalar $f'$ with two different $\gamma_{NR}$~(18.3 or 14.7)  and $\gamma_{RL}$~(7.2 or 5.0) are compared with those of candidates in PDG. The $\phi_3(1850)$ are supposed to dominantly decay to the two observed channels. The unit is MeV. }
\label{fp results}
\end{table*}

\begin{table*}[t]
\centering
\renewcommand\arraystretch{1.3}
\tabcolsep=0.3cm
\begin{tabular}{cllcccc}
\toprule[1.6pt]
$J^{PC}(n^{2s+1}L_J)$ & Decay & $\Gamma_{\text{exp}}$\cite{ParticleDataGroup:2024cfk} & $\Gamma_{NL}(\gamma_{NR}=18.3/14.7)$ & $\Gamma_{RL}(\gamma_{RL}=7.2/5.0)$  \\
\toprule[0.8pt]
$1^{--}(2^3S_1 )$ & $\omega(1420) \to \rho\pi$ &  $202.7\pm 133.1$ & 276/178 & 500/239 \\
\cline{1-5}
$2^{++}(1^3P_2)$ & $f_2(1270) \to \pi\pi $  &  155.8$\pm 5.5$ & 231/149 & 155/74 \\
$2^{++}(1^3P_2)$ & $f_2(1270) \to K\bar{K} $   & $8.5\pm 0.8$ & 2/1 & 3/2 \\
\cline{1-5}
$4^{++}(1^3F_4)$ & $f_4(2050) \to \omega\omega$ &  $40.3\pm 4.7$ & 165/106 & 349/167\\
$4^{++}(1^3F_4)$ & $f_4(2050) \to \pi\pi$ &  $35.4\pm 3.8$ & 92/59 & 59/28 \\
$4^{++}(1^3F_4)$ & $f_4(2050) \to  K\bar{K}$ &  $1.6\pm 0.7$ & 1/0.6 & 3/1.6 \\
\toprule[1.6pt]
\end{tabular}
\caption{The strong decay widths of isoscalar $f$ mesons with two different $\gamma_{NR}$~(18.3 or 14.7)  and $\gamma_{RL}$~(7.2 or 5.0) are compared with those of candidates in PDG. The unit is MeV. }
\label{f results}
\end{table*}

\begin{table*}[t]
\centering
\renewcommand\arraystretch{1.3}
\tabcolsep=0.3cm
\begin{tabular}{cllcccc}
\toprule[1.6pt]
$J^{PC}(n^{2s+1}L_J)$ & Decay & $\Gamma_{\text{exp}}$\cite{ParticleDataGroup:2024cfk} & $\Gamma_{NL}(\gamma_{NR}=18.3/14.7)$ & $\Gamma_{RL}(\gamma_{RL}=7.2/5.0)$  \\
\toprule[0.8pt]
$1^{--}(1^3S_1)$ & $K^*(892) \to K\pi $  &  $51.4\pm 0.8$ & 50/32 & 59/28 \\
\cline{1-5}
$0^{++}(1^3P_0 )$ & $K_0^*(1430) \to K\pi $  &  $251\pm 80$ & 1076/393 & 714/342 \\
\cline{1-5}
$2^{++}(1^3P_2 )$ & $K_2^*(1430) \to K\pi $  &  $54.4\pm 3.0$ & 124/80 & 97/46 \\
$2^{++}(1^3P_2 )$ & $K_2^*(1430) \to K^*(892)\pi $   & $26.9\pm 2.1$ & 51/33 & 66/32 \\
$2^{++}(1^3P_2 )$ & $K_2^*(1430) \to K\rho $  &  $9.5\pm 1.0$ & 17/11 & 28/14 \\
$2^{++}(1^3P_2 )$ & $K_2^*(1430) \to K\omega $  &  $3.2\pm 0.9$ & 5/3 & 8/4 \\
\cline{1-5}
$1^{--}(1^3D_1) $ & $K^*(1680) \to K\pi $  &  $123.8\pm 43.4$ & 239/154 & 140/67 \\
$1^{--}(1^3D_1)$ & $K^*(1680) \to K\rho $  &  $100.5\pm 36.8$ & 125/80 & 132/63 \\
$1^{--}(1^3D_1)$ & $K^*(1680) \to K^*(892)\pi $  &  $95.7\pm 35.0$ & 127/82 & 111/53 \\
\cline{1-5}
$3^{--}(1^3D_3) $ & $K_3^*(1780) \to K\pi $  &  $30.3\pm 3.6$ & 84/54 & 63/30 \\
$3^{--}(1^3D_3) $ & $K_3^*(1780) \to K\rho $  &  $49.9\pm 15.4$ & 50/32 & 74/35 \\
$3^{--}(1^3D_3 )$ & $K_3^*(1780) \to K^*(892)\pi $  &  $32.2\pm 8.9$ & 66/43 & 78/37 \\
\cline{1-5}
$4^{++}(1^3F_4 )$ & $K_4^*(2045) \to K\pi $  &  $19.7\pm 3.3$ & 45/29 & 38/18 \\
\cline{1-5}
$5^{--}(1^3F_5 )$ & $K_5^*(2380) \to K\pi $  &  $11.0\pm 3.8$ & 32/21 & 25/12 \\
\toprule[1.6pt]
\end{tabular}
\caption{The strong decay widths of strange mesons $\left(u\bar{s}, d\bar{s}, s\bar{u}, s\bar{d}\right)$ with two different $\gamma_{NR}$~(18.3 or 14.7)  and $\gamma_{RL}$~(7.2 or 5.0) are compared with those of candidates in PDG. The unit is MeV. }
\label{us results}
\end{table*}

\section{Numerical analyses and discussions}\label{section3}

To compare the predictive power of the two versions of the QPC models, one must adopt a specific description of the wave functions for both the initial and final mesons. In the present work, we employ the Godfrey-Isgur (GI) model~\cite{Godfrey:1985xj}, which has achieved broad phenomenological success in meson spectroscopy across both light and heavy flavor sections, particularly for states below the open-flavor thresholds. 
We acknowledge  that this choice also introduces an inherent degree of model dependence into the analysis. The GI model is itself based on a relativized quark-model framework designed primarily for spectroscopy, whereas the decay amplitudes considered here are evaluated within two different versions of the QPC mechanism. In particular, when the GI wave functions are used in the relativistic QPC calculation, the resulting framework should not be regarded as a fully self-consistent relativistic treatment derived from a common dynamical scheme. Rather, it should be viewed as a phenomenological setup intended to isolate, as much as possible, the impact of relativistic effects in the decay operator and amplitude construction while keeping the underlying meson wave functions fixed. This strategy does not eliminate model uncertainties, but it does provide a controlled basis for comparing the non-relativistic and relativistic QPC formulations on equal footing.

At the same time, fixing the meson wave functions to those of a well-known spectroscopic model prevents the introduction of arbitrary flexibility. The necessary compromise, however, is that one cannot retune the wave functions to optimize agreement with experimental partial widths for individual channels. Therefore, the quantitative results reported below are intended to illustrate the relative behavior of the non-relativistic and relativistic QPC formulations within a shared phenomenological framework, rather than to serve as a definitive evaluation of their absolute precision.

The pair-creation strength parameters in the NRQPC and RQPC models, denoted by $\gamma_{\rm NR}$ and $\gamma_{\rm RL}$ respectively, are both free parameters, and there is no \emph{a priori} relation connecting them. Because the spatial wave functions are strictly fixed to those of the  GI's model, $\gamma$ represents the sole adjustable degree of freedom in each formulation. 
To facilitate a robust and direct comparison, we adopt two kinds of calibration schemes. In the first scheme, we determine the $\gamma_{\rm NR}$ and $\gamma_{\rm RL}$ individually by calibrating against the well-measured decay width of the $\rho$ meson, and subsequently use these values to predict the partial widths of the remaining states. In the second scheme, we determine the optimal $\gamma$ parameter through a global fit to the experimental decay widths of a selected set of well-established mesons. The fitted sample consists of light-meson states for which experimentally measured two-body partial widths are available in the Particle Data Group (PDG) compilation~\cite{ParticleDataGroup:2024cfk}.
The light unflavored scalar resonances, such as the $f_0$ and $a_0$ states, are not included in the present analysis. This is mainly because these states typically have large decay widths, which can lead to substantial overlap among neighboring resonances, or because their line shapes may be significantly distorted by nearby open thresholds within a relatively narrow mass region~\cite{Close:2002zu}. Throughout these calculations, the physical masses of the initial- and final-state mesons are also taken from the latest  compilation of PDG to ensure the correct evaluation of the phase space.

For channels involving strange-quark pair creation, we further assume the commonly used empirical suppression
\begin{align}
\gamma_{s\bar s}=\frac{\gamma_{u\bar u}}{\sqrt{3}},
\end{align}
which means the $s\bar{s}$ pair is much harder to be created from the vacuum than the $u\bar{u}$ and $d\bar{d}$
and apply the same prescription in both the non-relativistic and relativistic calculations. This assumption is introduced only to maintain consistency with standard phenomenological practice; it may itself constitute an additional source of model dependence in the predictions for strange-meson decay channels.

We have calculated the strong decay widths for a broad class of mesons, including light unflavored $n\bar n$, $s\bar s$, and strange $n\bar s$ states. The isoscalar mesons, denoted by $f$ and $f'$ as in the PDG table, are treated as mixtures of the nonstrange and strange flavor components. In the $(u\bar u,d\bar d,s\bar s)$ basis, the physical states are written as
\begin{align}
f' &= \frac{1}{\sqrt{2}}(u\bar u+d\bar d)\cos\alpha - s\bar s\sin\alpha, \\
f  &= \frac{1}{\sqrt{2}}(u\bar u+d\bar d)\sin\alpha + s\bar s\cos\alpha ,
\end{align}
where $\alpha$ is the mixing angle. In the present work, we adopt the ideal-mixing approximation, under which $f'\approx s\bar s$ and $f\approx (u\bar u+d\bar d)/\sqrt{2}$~\cite{Godfrey:1985xj}.

The $\rho(770)$ meson decays almost entirely into the $\pi\pi$ channel and its width is measured accurately. Consequently, using this state ensures that no uncertainties from coupled channels are introduced into the calibration. By determining the strength parameters solely from the $\rho$ meson, 
$ M_{\rho}=775.26~\mathrm{MeV}$  and $\Gamma_{\rho}=147.4~\mathrm{MeV}$,  we obtain: 
\begin{align}
\gamma_{\rm NR}=18.3, \qquad \gamma_{\rm RL}=7.2,
\end{align}
where the subscripts $NR$ and $RL$ represent the corresponding values in the non-relativistic version and the relativistic one. This notation is maintained throughout the remainder of the paper.

For the second calibration scheme, the optimal $\gamma$ parameters are extracted by minimizing the function
$\chi^2=\sum_n(\frac{\Gamma^{QPC}_n-\Gamma^{exp}_n}{\sigma_n})^2$
over the selected set of meson decays. Rather than using the purely experimental uncertainties, which are often statistically very small and would cause certain channels to disproportionately dominate the fit, we assign a uniform tolerance of $\sigma_n=10$ MeV to all channels. This uniform weighting effectively accounts for the inherent theoretical uncertainties typical of phenomenological quark models and ensures a balanced global fit. The resulting best-fit parameters are:
\begin{align}
\gamma_{\rm NR}=14.7\pm 1.1, \qquad \gamma_{\rm RL}=5.0\pm 0.4.
\end{align}

The strong decay widths of the selected light mesons, calculated with the universal $\gamma$ parameters determined above, are summarized in Tables~\ref{uu results}--\ref{us results}. For each entry, the results obtained with the two different $\gamma$ values are listed together and separated by a ``/'' symbol.
 A global inspection of these results indicates that, for most decay channels, the widths predicted by the non-relativistic and relativistic formulations are numerically similar. However, it must be noted that when the meson wave functions are strictly fixed to the GI model and a single, universal $\gamma$ is employed, the predicted widths generally agree with experimental measurements only at the order-of-magnitude level. This behavior corroborates the findings of Ref.~\cite{Wang:2022xxi}, which demonstrated that state-dependent $\gamma$ values are required to accurately reproduce the strong decay widths across the $\rho$ and $\omega$ families.

Furthermore, the extracted value of $\gamma_{\rm NR}=18.3$ from the single $\rho-\pi\pi$ channel calibration is notably larger than the typical values adopted in many non-relativistic QPC studies. Nevertheless, this large value is entirely consistent with the recognized systematic tendencies of the QPC model. Historically, QPC calculations tend to significantly underestimate the $\rho-\pi\pi$ decay width unless a heavily inflated pair-creation strength is imposed. For instance, in Ref.~\cite{Pang:2018gcn}, a non-relativistic QPC calculation using $\gamma=11.6$ yields $\Gamma_\rho = 68.9$ MeV, still falling well short of the experimental width. Similarly, Ref.~\cite{Ackleh:1996yt} obtained $\Gamma_\rho \approx 79$ MeV using simple harmonic oscillator wave functions~($\beta=0.397$ GeV) and a strength parameter equivalent to $\gamma= 0.506	\times\sqrt{96\pi}=8.8$. Consequently, forcing the model to fully reproduce the experimental $\rho$ decay width inevitably necessitates a highly elevated $\gamma_{NR}$ parameter.

Ultimately, these results imply that the underlying dynamics of light-meson decays are too complex to be fully captured by a single, universal pair-creation parameter. This conclusion is strongly supported by recent non-relativistic QPC analyses; for example, Ref.~\cite{Wang:2022xxi} showed that simultaneously reproducing the decay widths of the $\rho$, $\rho_3$, $\omega$ and $\omega_3$ families requires $\gamma$ to vary substantially, ranging from roughly 4.14 to 14.6 depending on the specific initial state. Therefore, the order-of-magnitude deviations observed in our results reflect the intrinsic limitations of the universal $\gamma$ assumption in the light sector.

Based on the present analysis of light-meson decays, several key properties of the QPC formulations can be summarized as follows:
\begin{enumerate}
\item First, it is evident that a single, universal pair-creation parameter is insufficient to accurately describe the partial decay widths of all light-meson states simultaneously. Crucially, our results demonstrate that this fundamental limitation cannot be remedied simply by incorporating relativistic effects.  As shown in Tables~\ref{uu results}--\ref{us results}, when $\gamma$ is calibrated to precisely fit the $\rho$ meson, the predictions for the $b_1(1235)$, $\pi_{2}(1670)$, $\omega(1420)$ and $K^*(892)$  are consistent with experimental data. However, sizable discrepancies emerge for other states, generally restricting the agreement to an order of magnitude. It should be noted that this conclusion is strictly tied to the use of GI model wave functions; whether an alternative potential model could naturally resolve these discrepancies remains an open question.
\item Second, despite deviations in the absolute decay widths, the branching fractions between different partial widths of the same state are reasonably well described by both QPC models. For instance, although the predicted absolute partial widths of the $a_2(1320)$ to the $\rho\pi$, $\pi\eta$ and $K\bar{K}$ 
  channels are smaller than experimental measurements, both the non-relativistic and relativistic models successfully capture the experimental branching ratio $\mathcal{B}(a_2(1320)\rightarrow \rho\pi):\mathcal{B}(a_2(1320)\rightarrow  \pi\eta):\mathcal{B}(a_2(1320)\rightarrow  K\bar{K})\approx 15:3:1$. Similarly, the ratio $\mathcal{B}(f'_2(1525)\rightarrow K\bar{K})/\mathcal{B}(f'_2(1525)\rightarrow  \eta\eta)\approx 8.4\pm 2.2$ 
 is well reproduced, with the non-relativistic and relativistic formulations yielding values of approximately 5.6 and 6.0, respectively.
\item Finally, while the predictive power of both formulations is comparable for unflavored mesons, the relativistic QPC model yields systematically better partial decay widths for the strange mesons (see Table~\ref{us results}). Using the global fit value $\gamma_{RL}=5.1$, the predicted $K^*(892)$ width is 28 MeV, roughly half of the experimental 51.4 MeV,  which parallels the suppression observed for the $\rho$ meson. With this single exception, the relativistic QPC model obtains highly consistent partial widths for the higher-mass strange mesons, including the $K_2^*(1430)$, $K^*(1680)$, $K_3^*(1780)$ , $K_4^*(2045)$ and $K_5^*(2380)$. Furthermore, while the predicted values for the $K^*(1680)$ are smaller than the central experimental values, they remain viable once the significant experimental uncertainties associated with this state are appropriately acknowledged.
\end{enumerate}

\begin{figure*}[tb]
\centering
\includegraphics[width=5cm]{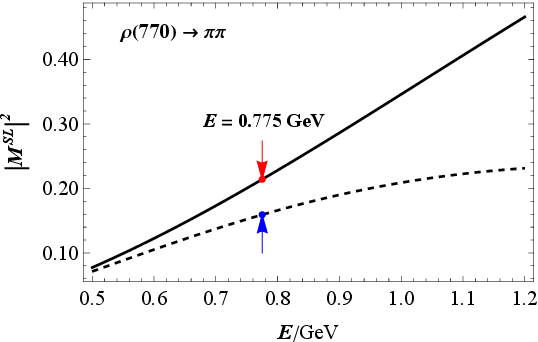}
\includegraphics[width=5cm]{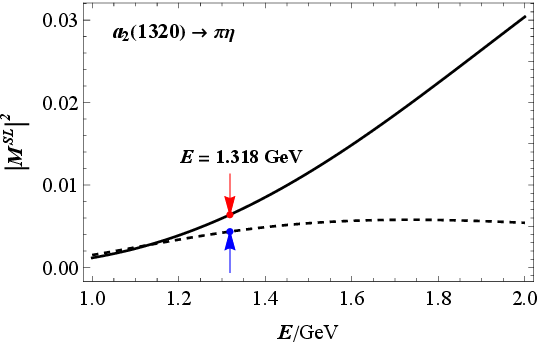}
\includegraphics[width=5cm]{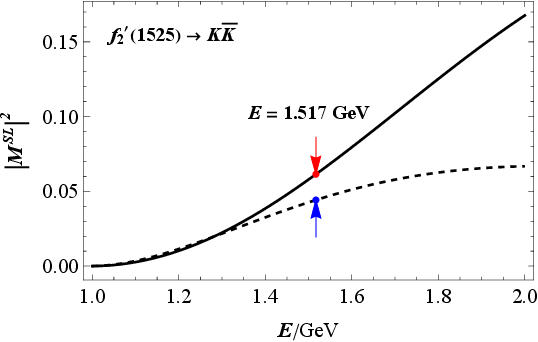}
\caption{The squared amplitudes $|M^{SL}|^2$ at the energy of initial mesons for $ \rho \to \pi\pi$, $ a_2(1320) \to \pi\eta$ and $ f_2'(1525) \to K\bar{K}$.  Solid curves are obtained from the non-relativistic QPC model with $\gamma_{NR}=14.7$ , while dashed curves represent the relativistic QPC results with $\gamma_{RL}=5.0$.}\label{amplocation}
\end{figure*}

\begin{figure*}[tb]
\centering
\includegraphics[width=5cm]{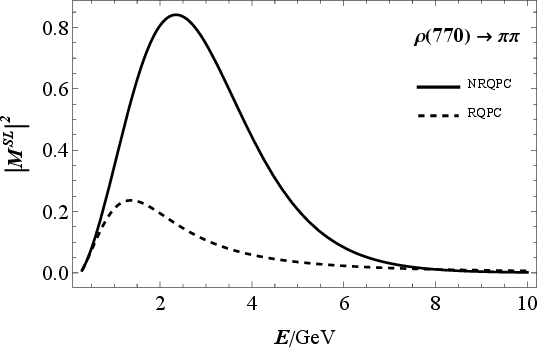}
\includegraphics[width=5cm]{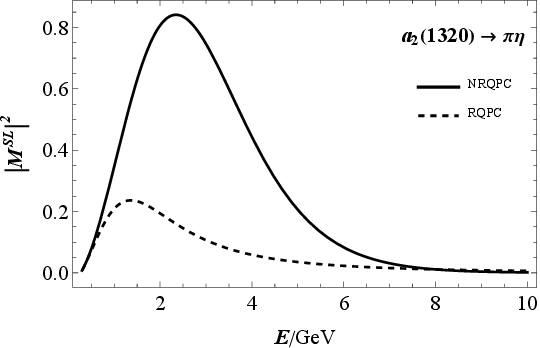}
\includegraphics[width=5cm]{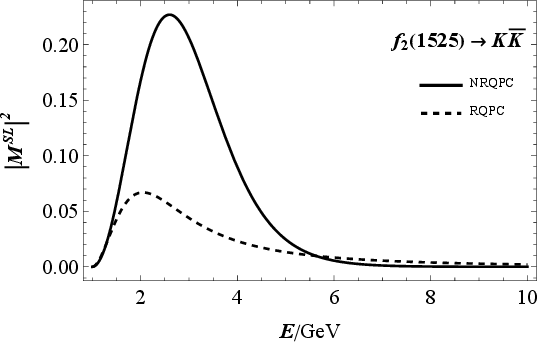}
\caption{Comparison of the squared amplitudes $|M^{SL}|^2$ of non-relativistic and relativistic ones in a larger off-shell energy range for three channels with the same parameters used in Fig.~\ref{amplocation}.  }\label{ampcompare}
\end{figure*}

Taken together, these observations suggest that, while the non-relativistic QPC model appears to provide an adequate qualitative description for low-lying states and low partial waves, relativistic effects may become increasingly important for highly excited mesons and for decay modes involving large momenta and high orbital angular momentum.

Although the integrated decay widths obtained in the non-relativistic and relativistic frameworks are numerically similar for most selected channels, the corresponding partial-wave scattering amplitudes $M^{SL}(E)$ can display rather different energy dependence. Such kind of energy dependence are not paid attention to in the isolated calculation of decay width, because  the three-momentum of the final-state mesons is fixed by the parent hadron mass $m_A$, which is usually set as
\begin{align}
q=\left.\sqrt{(E^2-(m_B+m_C)^2)(E^2-(m_B-m_C)^2)}/2E\right|_{E=m_A}.
\end{align}
When the QPC model is used in unquenched potential models to describe the off-mass-shell coupling vertex functions between a bare state and the relevant scattering channels (see, e.g., Refs.~\cite{Heikkila:1983wd,Geiger:1996re} and the subsequent literature), the energy-dependence behavior could influence the results. In such applications, the self-energy function is often treated through a dispersion relation, for example in a relativistic form,
\begin{equation}
\mathrm{Re}\,\Pi(s)
=-\frac{1}{\pi}\int_{s_{th}}^{\infty}
\frac{\mathrm{Im}\,\Pi(s')}{s-s'}\,ds' ,
\label{eq:selfenergy}
\end{equation}
where the Mandelstam variable is $s=E^2$, and $s_{th}$ denotes the threshold value. When $\mathrm{Im}\,\Pi(s')$ is modeled within the QPC framework, which is usually written down as $\mathrm{Im}\,\Pi(s)=\rho(s)\left|M^{SL}(s)\right|^2$ with $\rho(s)$ representing the kinematic factor proportional to $q(s)/\sqrt{s}$. $\mathrm{Re}\,\Pi(s)$ may be regarded, at least at the phenomenological level, as contributing to the mass shift of the physical state relative to the bare mass.

As illustrative examples, the quantities $|M^{SL}(E)|^2$ for $ \rho \to \pi\pi$, $ a_2(1320) \to \pi\eta$ and $ f_2'(1525) \to K\bar{K}$ are shown in Fig.~\ref{amplocation}. When the variable $E$ is fixed at the mass of parent meson, $|M^{SL}|^2$ is relevant for the partial decay width. One could find from Fig.~\ref{amplocation} that the related values in the two formulations are relatively close, which means both frameworks lead to  similar predictions for the decay width. As the center-of-mass energy increases, however, the differences between the corresponding squared amplitudes become progressively more pronounced. In the non-relativistic version, the amplitude rises to a comparatively large maximum before eventually decreasing and approaching zero at energies of order $10$ GeV. By contrast, in the relativistic version the variation is considerably more moderate, as indicated by the dashed curve in Fig.~\ref{ampcompare}. 
This behavior suggests that, even when the integrated widths are similar near the physical mass, the relativistic formulation exhibits a substantially milder high-energy dependence than its non-relativistic counterpart.

Since the dispersion integral of Eq.~(\ref{eq:selfenergy}) extends from threshold to infinity, the resulting mass-shift $\mathrm{Re}\,\Pi(s)$  can in principle be sensitive to the behavior of the transition amplitude over a broad energy range, including regions far above the physical decay point. In conventional non-relativistic QPC-based analyses, such effects have often been associated with rather sizable mass shifts, sometimes of the order of several hundred MeV. As indicated in Fig.~\ref{ampcompare}, however, the amplitude of relativistic framework displays a comparatively milder energy dependence at high energies. If such a trend remains relevant in self-energy applications, the relativistic formulation may reduce the relative importance of the high-energy region in the dispersion integral.

From this perspective, the slower falloff of $|M^{SL}|^2$ in the non-relativistic QPC model could enhance high-momentum contributions and thereby increase the sensitivity of the inferred mass shifts to the ultraviolet behavior of the model. By contrast, the relativistic formulation may suppress high-energy strength more efficiently. It means that including Lorentz-boost effects reduces the contribution from high energy region. 
Such a relativistic treatment could improve the convergence properties of the self-energy integral and lead to a more stable phenomenological determination of the physical mass. At the same time, it should be kept in mind that such conclusions are necessarily model dependent and may also be influenced by the treatment of subtraction constants, form factors, and wave-function parameters.

\section{summary}\label{section4}

In this work, we present a comparative study of the strong decays of light and strange mesons in both the non-relativistic and relativistic QPC frameworks. The relativistic formulation incorporates meson wave functions from the GI's model, exact kinematical phase space, and Wigner rotations of quark spins, and thus provides a more consistent treatment of decays involving sizable internal momenta. Partial-wave decay widths are evaluated for $n\bar n$, $n\bar s$, and $s\bar s$ mesons, including both low-lying and radially excited states such as the $K$, $K^*$, and isoscalar $f$ and $f'$ families.

We find that, for many of the light-meson decay channels considered in this work, the total widths predicted by the relativistic and non-relativistic QPC models are numerically similar, suggesting that relativistic effects do not produce substantial changes in the overall decay rates of these states. At the same time, it appears difficult to describe all total widths of light unflavored mesons accurately with a single universal pair-creation parameter, whereas the corresponding widths of strange mesons can be accommodated more naturally under the assumption of a universal $\gamma$. The overall branching patterns are reasonably reproduced in both frameworks. Whether the relativistic QPC formulation can provide a systematically improved description remains to be clarified through more extensive studies of excited-meson decays over a wider mass range, together with further experimental input. Nevertheless,  the energy dependence of the transition amplitudes is affected more noticeably: the relativistic formulation generally exhibits a stronger suppression at larger available energies.  These properties  suggest that relativistic effects may become particularly relevant in applications sensitive to the energy dependence of the QPC transition amplitudes, such as the coupled-channel analysis or quenched quark models with the coupling function introduced with the QPC dynamics.

\textit{Acknowledgement}: Helpful discussions with Cheng-Qun Pang are appreciated. This work is supported by China National Natural Science Foundation under contract No. 12375132, No. 11975075.


\appendix

\section{Lorentz-invariant two-particle state}\label{appendixA}

We adopt the canonical Lorentz boost  as the basic building block. we denote a four-momentum by $p=(p^{0},\mathbf{p})$, and use $l_{c}(p)$ to represent the canonical Lorentz boost from the original inertial frame to a frame moving with velocity $\mathbf{v}=-\mathbf{p}/p^{0}$.
If \(q=(q^0,\mathbf q)\) is a four-vector in some original inertial frame, its image under the canonical boost to the moving frame is written compactly as 
\begin{equation}
l_c(p)\,q = l_c(p)\begin{pmatrix} q^0 \\ \mathbf q \end{pmatrix}
= \begin{pmatrix}
\displaystyle \frac{p^0 q^0 + \mathbf p\!\cdot\!\mathbf q}{W} \\[6pt]
\displaystyle \mathbf q + \frac{\mathbf p}{W}\left( \frac{\mathbf p\!\cdot\!\mathbf q}{p^0+W} + q^0 \right)
\end{pmatrix},
\end{equation}
where $W=\sqrt{p\!\cdot\!p}$.
The one-particle momentum--spin state with mass \(\mu\), spin \(s\) and third-component \(m\) is defined by acting with the unitary representation of this boost on the rest-frame spin eigenstate and including the usual relativistic normalization factor:
\begin{equation}
|\mathbf p, s m\rangle = U\bigl(l_c(p)\bigr)\,| \mathbf 0, s m\rangle \sqrt{\frac{\mu}{\varepsilon(\mathbf p)}},
\end{equation}
with \(\varepsilon(\mathbf p)=\sqrt{\mathbf p^{\,2}+\mu^2}\), $\sqrt{\frac{\mu}{\varepsilon(\mathbf p)}}$ is the normalization coefficient. Under a general Lorentz transformation \(a\) the one-particle basis transforms with an accompanying Wigner rotation \(r_c(a,p)\); the representation matrix element reads \cite{Martin:1970hmp}
\begin{equation}
U(a)|\mathbf p, s m \rangle = \sum_{m'} \, |a\mathbf p, s m'\rangle\, \mathscr{D}^{(s)}_{m' m}\bigl(r_c(a,p)\bigr) \sqrt{\frac{\varepsilon(a\mathbf p)}{\varepsilon(\mathbf p)}},
\end{equation}
where $r_c(a,p)=l_c^{-1}(ap)al_c(p)$ is the Wigner rotation, $\mathscr{D}^{(s)}_{m' m}\bigl(r_c(a,p)\bigr)$ is the standard matrix representation of the rotation $r_c(a,p)$.

Based on the definition of the single-particle state, we construct two-particle~(denoted as particle ``1" and ``2") kinematics by first specifying the center-of-mass (c.m.) variables \cite{Fuda:2012xd}. In the c.m.\ frame the two on-shell four-momenta are
\[
k_1=(\varepsilon_1(\mathbf k),\mathbf k), ~~ k_2=(\varepsilon_2(-\mathbf k),-\mathbf k),
 \varepsilon_i(\mathbf k)=\sqrt{\mathbf k^{\,2}+\mu_i^2},
\]
where $\mathbf k$ is the three-momentum of particle 1 in the c.m. frame of the two-particle system. Boosting this two-particle configuration to a general total four-momentum \(p=(p^0,\mathbf p)\) by the canonical boost \(l_c(p)\) gives the lab-frame single-particle four-momenta \(p_i=l_c(p)k_i\) with \(p_1+p_2=p\). One may then define a boosted two-particle canonical state by
\begin{align}
| \mathbf {pk}, s_1 m_1 s_2 m_2 \rangle &= U\bigl(l_c(p)\bigr)|\mathbf k, s_1 m_1\rangle\otimes |-\mathbf k, s_2 m_2\rangle\sqrt{\frac{W(\mathbf{k})}{E(\mathbf{p},\mathbf{k})}},\notag\\
W(\mathbf{k})&=\varepsilon_1(\mathbf{k})+\varepsilon_2(-\mathbf{k}),\notag\\
E(\mathbf{p},\mathbf{k})&=\sqrt{\mathbf{p}^2+W(\mathbf{k})^2},
\end{align}
and the two-particle state mentioned above could be expressed as the combination of the direct product of two single-particle states multiplied with extra factors of the matrix representations for the Wigner rotations of the two particles
\begin{align}
|\mathbf{pk}, s_1 s_2 m_1 m_2\rangle 
=& \sum_{m_1' m_2'}
|\mathbf{p}_1, s_1 m_1'\rangle \otimes |\mathbf{p}_2, s_2 m_2'\rangle\notag\\
  & \mathscr{D}^{s_1}_{m_1' m_1}\!\left[r_c(l_c(p), k_1)\right]
  \mathscr{D}^{s_2}_{m_2' m_2}\!\left[r_c(l_c(p), k_2)\right]\notag\\
  &  
\left(
\frac{\varepsilon_1(\mathbf{p}_1)\,
      \varepsilon_2(\mathbf{p}_2)\,
      W(\mathbf{k})}
     {\varepsilon_1(\mathbf{k})\, 
      \varepsilon_2(-\mathbf{k})\,
      E(\mathbf{p},\mathbf{k})}
\right)^{1/2}.
\end{align}

Similarly, one can perform the partial wave decomposition and couple the orbital angular momentum $l$ and total spin $s$ to form the total angular momentum $j$ in the center-of-mass (c.m.) frame. Then, by boosting to a general frame, one can express the two-particle state in the angular momentum representation as
\begin{widetext}
\begin{align}
|\mathbf{p}kl s j m\rangle
&= \sum_{m_l, m_s, m_1, m_2}
\int d\Omega_\mathbf{k}\, 
Y_l^{m_l}(\hat{\mathbf{k}})\,
|\mathbf{pk}, s_1 s_2 m_1 m_2\rangle
\langle s_1 s_2 m_1 m_2| s m_s\rangle
\langle l s m_l m_s| j m\rangle\notag\\
&= 
\sum_{m_l, m_s, m_1, m_2, m_1', m_2'}
\int d\Omega_\mathbf{k}\, 
|\mathbf{p}_1, s_1 m_1'\rangle
\otimes 
|\mathbf{p}_2, s_2 m_2'\rangle\,
Y_l^{m_l}(\hat{\mathbf{k}})\,
\langle s_1 s_2 m_1 m_2| s m_s\rangle
\langle l s m_l m_s| j m\rangle
\nonumber \\[4pt]
&\quad \times 
\mathscr{D}^{s_1}_{m_1' m_1}
\!\left[r_c(l_c(p), k_1)\right]
\mathscr{D}^{s_2}_{m_2' m_2}
\!\left[r_c(l_c(p), k_2)\right]
\left(
\frac{
\varepsilon_1(\mathbf{p}_1)\, 
\varepsilon_2(\mathbf{p}_2)\, 
W(\mathbf{k})
}{
\varepsilon_1(\mathbf{k})\, 
\varepsilon_2(-\mathbf{k})\, 
E(\mathbf{p}, \mathbf{k})
}
\right)^{1/2}.
\end{align}
\end{widetext}
where $k$ is the magnitude of the three-momentum $\mathbf{k}$ and $\hat{k}$ is direction of $\mathbf{k}$.

\section{The table of mass spectrum of selected states in the PDG}

The masses and widths of the meson states used in this work are listed in Table \ref{spectrum}.

\begin{table*}[t]
\centering
\renewcommand\arraystretch{1.3}
\tabcolsep=0.1cm
\begin{tabular}{c|cccc}
\toprule[1.56pt]
System & States & PDG states & PDG masses & PDG width   \\
\hline
\multirow{11}{*}{$u\bar{d}, \frac{1}{\sqrt{2}}(u\bar{u}-d\bar{d}), d\bar{u}$}
& $1^1S_0$ & $\pi$ & 134.9/139.5 & ...   \\
& $1^3S_1$ & $\rho(770)$ & 775.26$\pm$0.23 & 147.4$\pm$0.8   \\
& $2^3S_1$ & $\rho(1450)$ & 1465$\pm$25 & 400$\pm$60    \\
& $1^3D_1$ & $\rho(1700)$ & 1720$\pm$20 & 250$\pm$100  \\
& $1^1P_1$ & $b_1(1235)$ & 1229.5$\pm$3.2 & 142$\pm$9    \\
& $1^3P_1$ & $a_1(1260)$ & 1230$\pm$40 & 380$\pm$80   \\
& $1^3P_2$ & $a_2(1320)$ & 1318.2$\pm$0.6 & 107$\pm$5   \\
& $2^3P_2$ & $a_2(1700)$ & 1706$\pm$14 & 380$_{-50}^{+60}$   \\
& $1^1D_2$ & $\pi_2(1670)$ & $1670.6_{-1.2}^{+2.9}$ & $258_{-9}^{+8}$   \\
& $1^3D_3$ & $\rho_3(1690)$ & 1688.8$\pm$2.1 & 161$\pm$10   \\
\hline
\multirow{8}{*}{$u\bar{s}$, $d\bar{s}$, $s\bar{u}$, $s\bar{d}$}
& $1^1S_0$ & $K$ & 497.6/493.7 & ...    \\
& $1^3S_1$ & $K^*(892)$ & 891.67$\pm$0.26 & 51.4$\pm$0.8    \\
& $1^3P_2$ & $K_2^*(1430)$ & 1427.3$\pm$1.5 & 100.0$\pm$2.2   \\
& $1^3D_1$ & $K^*(1680)$ & 1718$\pm$18 & 320$\pm$110    \\
& $1^3D_3$ & $K_3^*(1780)$ & 1779$\pm$8 & 161$\pm$17    \\
& $1^3F_4$ & $K^*(2045)$ & $2048_{-9}^{+8}$ & $199_{-19}^{+27}$    \\
& $1^3G_5$ & $K^*(2380)$ & 2382$\pm$24 & 180$\pm$50   \\
\hline
\multirow{4}{*}{$f'$}
& $1^1S_0$ & $\eta$ & 547.862$\pm$0.017 & 1.31$\pm$0.05    \\
& $1^3S_1$ & $\phi(1020)$ & 1019.460$\pm$0.016 & 4.249$\pm$0.013    \\
& $1^3P_2$ & $f_2'(1525)$ & 1517.3$\pm$2.4 & $72_{-6}^{+7}$   \\
& $1^3D_3$ & $\phi_3(1850)$ & 1854$\pm$7 & $87_{-23}^{+28}$    \\
\hline
\multirow{5}{*}{$f$}
& $1^1S_0$ & $\eta'(958)$ & 957.78$\pm$0.06 & 0.188$\pm$0.006   \\
& $1^3S_1$ & $\omega(782)$ & 782.66$\pm$0.13 & 8.68$\pm$0.13   \\
& $2^3S_1$ & $\omega(1420)$ & 1410$\pm$60 & 290$\pm$190   \\
& $1^3P_2$ & $f_2(1270)$ & 1275.4$\pm$0.8 & $185.8_{-2.1}^{+2.8}$    \\
& $1^3F_4$ & $f_4(2050)$ & 2018$\pm$11 & 237$\pm 18$    \\
\toprule[1.6pt]
\end{tabular}
\caption{The mass spectrum of quarkonium with those of candidates in PDG~\cite{ParticleDataGroup:2024cfk}. The unit is MeV. }
\label{spectrum}
\end{table*}
\


\clearpage
%
\bibliographystyle{apsrev4-2}
\bibliography{Ref}

\end{document}